\documentclass{appolb}
\usepackage{graphicx}
\usepackage{graphicx}
\usepackage{epsfig}
\usepackage{color}
\usepackage{longtable}
\usepackage{hyperref}
\usepackage{latexsym}
\usepackage{amsmath}
\usepackage{amssymb}
\usepackage{dsfont}
\usepackage{verbatim}
\usepackage{bm}

\newcommand{\unitmatrix}{\mathds{1}}

\begin{document}
\title{Nucleon excited states on the lattice
\thanks{Presented at ``Excited QCD 2013'', Bjelasnica Mountain, Sarajevo}%
}
\author{C.B. Lang, \underline{V. Verduci}
\address{Institut f\"ur Physik, Universit\"at Graz, \\A--8010 Graz, Austria}}
\maketitle
\begin{abstract}
We study the pion-nucleon system in $s$-wave in the framework of lattice QCD in order
to gain new information on the nucleon excited states. We perform simulations for
$n_f =2$ mass degenerate light quarks at a pion mass of 266 MeV. The
results show that including the two-particle states drastically changes the energy
levels. The variational analysis and the distillation approach play an
important role in the extraction of the energy levels. The phase shift analysis
allows to extract information on the resonance nature of the observed states.
\end{abstract}
  
\section{Introduction}
Almost all the hadrons that constitute the QCD spectrum are unstable under strong
interactions. Lattice QCD calculation have been traditionally treating these states
as stable states, without taking into account their resonant nature. Only recent
studies have made exploratory steps in this direction, successfully studying mesonic
resonances
\cite{Lang:2011mn,Aoki:2011yj,Feng:2010es,Lang:2012sv,Mohler:2012na,Pelissier:2012pi,Cohen:2009zk,Dudek:2012xn}.

We study for the first time the coupled pion-nucleon system explicitly  including
the two particles in our simulations \cite{Lang:2012db}. This work is motivated by
the fact that lattice hadron spectroscopy does not satisfactory reproduce the
negative parity sector of the nucleon states. The physical spectrum consists of two
resonances $N^*(1535)$ and $N^*(1650)$. So far lattice simulations
\cite{Mahbub:2009cf,Dudek:2010wm,Bulava:2010yg,Engel:2010my,Engel:2012qp}
have measured in this channel two low-lying states that are assigned to the two resonances,
even though the lower measured state lies below the physical value of $N^*(1535)$
\cite{Engel:2013ig}. All these simulations considered only $3$-quark interpolators that
should in principle couple to meson-baryon states via dynamical quark loops. However
this coupling seems to be weak and meson-baryon interpolators have to be
explicitly included in the set of operators in order to achieve a complete study of
these resonances.

The negative parity resonances of the nucleon couple to $N\pi$ in $s$-wave, but it is not the
only decay channel: $N^*(1535) \rightarrow N\eta$,  $N^*(1650)\rightarrow N\eta,
\Lambda K$ \cite{Beringer:1900zz}. However, we work at an unphysical pion mass
($m_{\pi}$ = 266 MeV) that prevents this channel from being in the influence region
of the two resonances. We therefore simulate on our lattice the coupled channel of
a 3-quark nucleon together with the (4+1)-quark $N\pi$ system in the rest frame. The
results presented here have already been published in \cite{Lang:2012db}.

\section{Tools and setup}

\subsection{Variational analysis}
The energy levels of the nucleon and the $N\pi$ system are determined using the
variational method \cite{Michael:1985ne,Luscher:1985,Luscher:1990ck}. We measure the
Euclidean cross-correlation matrix $C(t)$ between different interpolators $O_i(t)$
\begin{equation}
 C_{ij}(t)=\langle O_i(t) \,\bar{O}_j(0)\rangle =
 \sum_n\langle O_i(t)|n\rangle e^{-E_n t}\langle n|\bar{O}_j(0)\rangle\,,
\end{equation}
and then solve the generalized eigenvalue problem
\begin{equation}
C(t)\vec u_n(t)=\lambda_n(t)C(t_0)\vec u_n(t)
\end{equation}
to disentangle the eigenstates with the eigenvalues
$\lambda_n(t,t_0)\sim \e^{-E_n (t-t_0)}$. The energy values of the
eigenstates are determined by exponential fits. The fit range is indicated by a plateau-like behavior of the effective masses 
$E_n(t)=\log [{\lambda_n(t)}/{\lambda_n(t+1)}]$.

\subsection{Distillation}
The evaluation of the correlation matrix in the case of (4+1) quarks turns out to be
hardly accessible to traditional techniques, due to the large amount of different
diagrams involved. The distillation method \cite{Peardon:2009gh} allows to evaluate
partially disconnected diagrams within an affordable amount of computer time.  The
quark sources are smeared using a truncated expansion of the 3D Laplacian operator 
\begin{equation}
q(x)  \rightarrow {S(x,x^{'})} q(x^{'}) =  
{\sum_{i=1}^{N_v}\, v^i(x)v^{i^\dagger}(x^{'})}\,q(x^{'})\,.
\end{equation}
The correlation function for the 3-quark nucleon operator reads
\begin{equation}
C(t_{snk},t_{src})
=\phi_{t_{snk}}(i,j,k) \, \, \, \tau(i,i')
\tau(j,j')
\tau(k,k') \,\,\,
\phi_{t_{src}}^{\dagger}(i',j',k')
\end{equation}
where the perambulators $\tau(n,m)$ are quark propagators from source eigenvector $v^m$
to sink $v^n$. The functions $\phi$ include all the information on the 
Dirac structure of the specific interpolator
\begin{equation}
\phi_{t_{snk}}(i,j,k) = 
\;\sum_{\vec x}\epsilon_{abc} D \,
v^{i}_a(\vec x)
v^{j}_b(\vec x)
v^{k}_c(\vec x)\,,
\end{equation}
where D carries all the Dirac indices.

\subsection{Interpolators}
The set of interpolators has to be as complete as possible in order to reliably extract the spectrum. 
We use the nucleon interpolator
\begin{equation}
N_\pm^{(i)}(\vec p=0)=\sum_{\vec x}\epsilon_{abc}\,  
P_\pm\,\Gamma_1^{(i)}\, u_a(\vec x)\, 
u_b^T{(\vec x)}\, \Gamma_2^{(i)}\, d_c{(\vec x)} \,,
\end{equation}
where
$(\Gamma_1,\Gamma_2)=(\unitmatrix,C\gamma_5),(\gamma_5,C),(i\unitmatrix,C\gamma_t\gamma_5)$
and each quark source is smeared  combining  $N_v=32$ and $64$ eigenvectors. For the $N\pi$
system we use
\begin{equation}
N\pi(\vec p=0)=\gamma_5 N_+(\vec p=0)\pi(\vec p=0)\,,
\end{equation}
and we project to isospin $1/2$: $\,O_{N\pi} = p \pi^0 + \sqrt{2}\, n \pi^+$ with
\begin{equation}
\pi^0(\vec 0)=\frac{1}{\sqrt{2}}\sum_{\vec x} \{\bar{u}_a(\vec x) \gamma_5 u_a(\vec x)
-\bar{d}_a(\vec x) \gamma_5 d_a(\vec x)\}\,,\;\; \pi^+(\vec 0)=\sum_{\vec x} \bar{d}_a(\vec x) \gamma_5 u_a(\vec x)
\,.
\end{equation}

\subsection{Interpretation of the energy levels}
Once the energy levels are computed, one can relate the measured spectrum 
of the coupled system to the 
physical resonances. In the elastic region L{\"u}scher's formula gives a relation
between the discrete energy levels (of the rest frame system, for a
discussion of meson-baryon systems in moving frames see \cite{Gockeler:2012yj}) and the phase shift in the continuum
\cite{Luscher:1985,Luscher:1990},
\begin{equation}\label{zeta}
\tan \delta(q)=\frac{\pi^{3/2} q}{\mathcal{Z}_{00}(1;q^2)}\,,
\end{equation}
where the generalized zeta function
$Z_{lm}$ is given in \cite{Luscher:1990} and 
\begin{equation}\label{eq:pstar_q2}
q=p^*\frac{L}{2\pi}\,,\;\; p^{*2}=\frac{[s-(m_N+m_\pi)^2][s-(m_N-m_\pi)^2]}{4s}
\end{equation}
with $s=(E_n)^2$. Given some phase shift model, eq.~\eqref{zeta} can be numerically
inverted to obtain  the expected energy levels for the two interacting particle
system (Fig. \ref{Fig:F2H}, rhs). For a different method to predict the expected energy levels in finite volume see e.g.\cite{meissner}.

\section{Results}
We use 280 configurations generated for two flavors of mass-degenerate light quarks and
a tree level improved Wilson-Clover action. $m_{\pi}=266$ MeV, $a = 0.12$ fm,
$V=16^3 \times 32$ \cite{Hasenfratz:2008}.

\subsection{One particle sector}
\begin{figure}[htb]
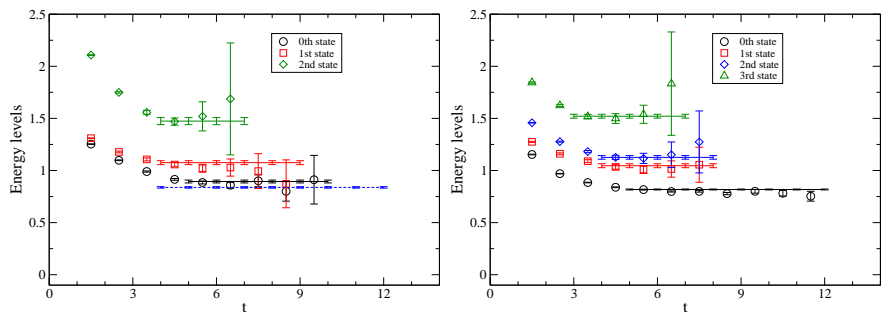

\centerline{
\includegraphics[width=0.45\linewidth,clip]{masses_negative_110000110.eps}
\includegraphics[width=0.45\linewidth,clip]{masses_negative_110111110.eps}
}
\caption{Effective mass values for $N_-$. Left: Results for 3 quark interpolators.
The blue line denotes the non-interacting $N\pi$ state. Right: Results for the $N_-$
and $N_+\pi$ coupled system.}
\label{Fig:masses_negative}
\end{figure}

Using a set of 3-quark interpolators we reproduce the usual observed
spectrum \cite{Engel:2013ig}. In the positive sector we observe the
nucleon ground state at $m_N=1068(6)$ MeV and another state that lies far
above the physical Roper. In the negative sector we observe two nearby levels,
the lowest lying below $N^*(1535)$ (Fig. \ref{Fig:masses_negative},
lhs).

\subsection{Coupled $N$ and $N\pi$ system}
First we compute the energy level for the two particles propagating
independently (i.e., the threshold) 
and we find that it is overlapping with the first of the
two levels measured in the single-particle approach (Fig.
\ref{Fig:masses_negative}, lhs). When $O_{N\pi}$ is included a new
energy level appears and the effective energy levels of the $N\pi$
system show less fluctuations compared to the 3-quark case. The
lowest level now lies slightly below the $N\pi$ threshold, a feature
typical for attractive $s$-wave and a finite volume artifact. The
next-higher two levels now lie approximately 130 MeV above the physical
resonance positions of $N^*(1535)$ and $N^*(1650)$, similar to the
shift of the nucleon ground state for this value of $m_\pi$.

\subsection{Phase shift analysis}
\begin{figure}[htb]
\centerline{
\includegraphics[height=5cm,clip]{levels_comparison.eps}\hfill
\includegraphics[height=5cm,clip]{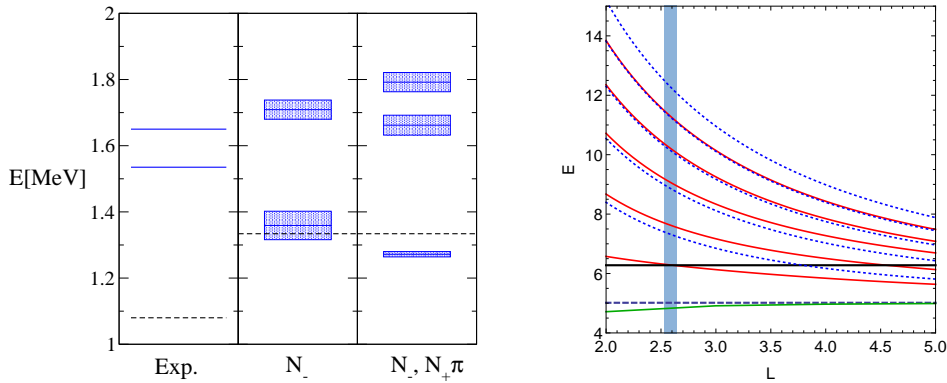}
}
\caption{Lhs: Comparison between the experimental masses of the negative 
parity nucleon resonances, the energy spectrum obtained from the single particle
 analysis and the results from the coupled $N$ and $N\pi$ system. 
 Rhs: Energy levels expected for the interacting $N\pi$ system 
 obtained inverting the L{\"u}scher formula and assuming a Breit-Wigner 
 parametrization for $N(1530)$.}
\label{Fig:F2H}
\end{figure}

A comparison with the expected energy levels obtained inverting
L{\"u}scher formula \eqref{zeta} and assuming a single elastic
resonance parameterization, show excellent agreement (Fig. 
\ref{Fig:F2H}, rhs). Assuming a Breit-Wigner shape for the first
resonance we can also extract the resonance mass: $m_R=1.678(99)$ GeV.

\section{Conclusions}
This study is intended to shed some light on the excited energy levels
of the nucleon spectrum, which  still represents an outstanding challenge for
lattice QCD. We find that including meson-baryon interpolators is indeed
needed for a reliable picture of the $N_-$ spectrum. The study of two
particle systems on the lattice improves our understanding of
LQCD and this work is a first step into that
direction.

\section{Acknowledgments}
We thank G. Engel, C. Gattringer,  L. Glozman, M. G\"ockeler, D. Mohler, C.
Morningstar, S. Prelovsek and A. Rusetsky for  valuable discussions. Special
thanks to Anna Hasenfratz for providing the dynamical configurations. The
calculations were performed on local clusters at UNI-IT at the University of
Graz.  V.V.~ has been supported by the Austrian Science Fund (FWF) under
Grant DK W1203-N16.


\end{document}